# The Importance of Variable Importance

Charles D. Coleman[a]

[a]*Charles D. Coleman is a Mathematical Statistician, U.S. Census Bureau, CENHQ 5H48C, Washington, DC, 20233 (e-mail: charles.d.coleman@census.gov).  Any views expressed are those of the author and not necessarily those of the U.S.  Census Bureau. The Census Bureau has reviewed this data product for unauthorized disclosure of confidential information and has approved the disclosure avoidance applied. (Approval ID: CBDRB-FY20-003)*

Variable importance is defined as a measure of each regressor's contribution to model fit. Using $R^2$ as the fit criterion in linear models leads to the Shapley value (LMG) and proportionate value (PMVD) as variable importance measures. Similar measures are defined for ensemble models, using random forests as the example. The properties of the LMG and PMVD are compared. Variable importance is proposed to assess regressors' practical effects or "oomph." The uses of variable importance in modelling, interventions and causal analysis are discussed.

## 1 Introduction

Measures of variable importance, sometimes called "relative importance," decompose a measure of the fit of a multivariate model into a sum of each regressor's contribution to fit. Variable importance is usually used to rank regressorss' contribution to the overall fit. This leads to the ability to select variables for intervention or analysis. These rankings have been used to select variables for analysis De Veaux (2007, p. 152) or intervention (for example, Fields (1993) and the references in Johnson and LeBreton (2004, pp. 240-242)). Generalizing "oomph," originally "a big change, important for the science" (Ziliak and McCloskey, 2004, p.531), to a continuous variable that measures "oomphiness" or substantive significance (Ziliak and McClosky, 2008, p. 27) enables the use of variable importance to measure it. Section 3 quantifies "oomph" as the amount of $R^2$ explained in a linear model, that is, variable importance as used in the present article for linear models. This enables variable importance to be used instead of or in addition to a $p$-value. However, like with a $p$-value, the analyst still has to provide a required cutoff or confidence interval. In effect, variable importance enables the substitution of substantive significance for statistical significance. Like any other statistic based on noisy data, variable importance is stochastic. Its uncertainty can be estimated using the bootstrap to produce confidence intervals.

Variable importance itself is not well-defined. Game theorists have been unable to produce a satisfactory definition for variance-based measures (Grömping, 2015, p. 149). Johnson and LeBreton (2004, p. 240) offer a working definition:

> *Relative importance*: The proportionate contribution each predictor makes to $R^2$, considering both its direct effect (i.e., its correlation with the criterion) and its effect when combined with the other variables in the regression equation. [Original italics]

The present article is less concerned with the exact definition in favor of description. This definition is generalized to use measures other than $R^2$ in general models by Chevan and Sunderland (1991) and in ensemble methods as described below. These measures are generally based on dispersion or a proxy. Variable importance can also be measured in terms of levels, such as van der Laan (2006) and Lundberg and Lee (2017). Ensemble models use MSE, a combination of both level and dispersion. The present article focuses on $R^2$ shares, equivalent to variance explained shares, for linear models and MSE reduction for ensemble models.

Wright (1921) first proposed a measure of variable importance in the context of causal path analysis. Dunlap and Cureton (1930) applied Wright's method to all possible causal paths in a linear model to obtain perhaps the first path-independent variable importance measure. Lindeman, Merenda and Gold (1984, p. 120) first proposed the Shapley (1953) value for linear models, though not by that name. Their statistic is called "LMG" in the context of variable importance. The LMG has been rediscovered many times, such as Kruskal (1987), Chevan and Sutherland (1991), Lipovetski and Conklin (2001) and Israeli (2007). Stufken (1992) was the first to demonstrate the equivalence of the LMG and Shapley value. The Proportionate Method Value Decomposition "PMVD" (Feldman, 2005) is the main competitor to the Shapley value, based on the proportionate value (Ortmann, 2000) with similar computations. Both of these have analogues in ensemble methods, the difference mainly lying in how ensemble methods choose variables for splits.

Perhaps the most widely taught variable importance decomposition method is Theil's (1971, pp. 168-181) "incremental contribution" approach. A search of the internet for "variable importance" is dominated by descriptions of it, even though Theil himself (Theil (1987) and Theil and Chung (1988)) described different methods based on information theory, an entirely different concept. Theil seems to have made no subsequent reference to incremental contributions. The widespread appearance of this method appears to be due to heavy use of Theil (1971) in econometrics classes. Few outside of econometrics and data science seem to have heard about variable importance. Section 3 discusses Theil's (1971) method and its flaws.

For surveys of applications of variable importance, see, for example, Johnson and LeBreton (2004, p. 242) and Grömping (2015, p. 142).

This article is concerned only with linear models and ensemble methods, focusing on random forests for the latter. Chevan and Sunderland (1991) propose a general method applicable to a larger class of models given a goodness-of-fit metric. Grömping (2015, p. 140) has references to specialized methods for some nonlinear models.

Section 2 describes variable importance measures for linear models and random forests, along with their properties. Section 3 describes using variable importance to measure "oomph." Section 4 discusses using variable importance in modelling. Section 5 shows how to combine different measures of variable importance with correlation analysis and analyst judgment to at least partially discern causal structure. Section 6 discusses interventions. Section 7 concludes this article.

## 2 Variable Importance Measures Defined

Variable importance, in general, can be defined as each variable's contribution to level or dispersion (Achen, 1982, pp. 68-77). Dispersion has generally been found more useful in the linear regression context while the variable importance measures considered in the present article for ensemble methods like random forests decompose reduction in MSE, a function of both level and dispersion. Decomposing $R^2$ in linear models is equivalent to decomposing the shares of variance explained by each variable. Rather than run full regressions to obtain $R^2$ from each, Grömping (2007a) describes an efficient variance-covariance matrix-based algorithm. Ensemble methods differ in two respects. First, the variable importance criterion is selected in advance, often implicitly, (Boulesteix et al., 2012). Second, the regressors' contributions to the metric need not sum to the total. In the case of random forests, the MSE reduction (related to Breiman's (2002) "Method 1" to measure variable importance) accounted for by each variable does not sum to total MSE reduction. The workaround is to apportion total MSE reduction in proportion to the individual variables' MSE reductions. Ensemble methods have additional methods to measure variable importance (Breiman (2002)). These are not covered in the present article. Ensemble methods are covered in Subsection 2.4.

Variable importance measures generally have the properties[1]

(1) *Proper decomposition*: the model fit statistic (e.g., $R^2$ or MSE reduction) is to be decomposed into shares, that is, the sum of all shares has to be the model fit statistic.

(2) *Non-negativity*: all shares have to be non-negative.

(3) *Inclusion*: a regressor $X_j$ with $\beta_j \neq 0$ should receive a nonzero share.

We focus first on the linear model. The LMG satisfies properties (1)-(3). Let the model be the usual linear model $\boldsymbol{y} = \boldsymbol{X\beta}$ with i.i.d. errors. Then, the LMG for each variable $X_i$ can be computed as[2]

$$LMG(X_j) = \sum_{M \in N \setminus \{X_j\}} \gamma_n(M)(R^2_{M \cup \{X_j\}} - R^2_M),$$

$$(1)$$

where $M$ is an $m$-variable subset of the $n$ variables $N$, the subscripts on $R^2$ indicate the regressors and

$$\gamma_n(M) = m!\,(n - m - 1)!/n!,$$

$$(2)$$

are the weights applied to each subset of variables. Equation (1) is a simplification of the unweighted average over all $n!$ permutations of the variables

$$LMG(X_j) = \frac{1}{n!} \sum_r \left( R^2_{P^r_j \cup \{X_j\}} - R^2_{P^r_j} \right),$$

$$(3)$$

where $r$ is a variable order and $P^r_j$ is the set of variables before $X_j$ in order $r$. Defining the sequential increment in $R^2$ caused by including $X_j$ after an order $r$ as

$$seqR^2(\{X_j\}|r) = R^2_{P^r_i \cup \{X_j\}} - R^2_{P^r_j}$$

$$(4)$$

allows us to rewrite equation (3) as

$$LMG(X_j) = \frac{1}{n!} \sum_r seqR^2(\{X_j\}|r).$$

(5)

Equation (5) will be useful for comparison with the PMVD below.

The LMG is the special case of hierarchical partitioning (Chevan and Sunderland, 1991) applied to the linear model with $R^2$ as the goodness-of-fit criterion.

The PMVD is calculated similarly to the LMG, replacing the equal weights in equation (3) with data-dependent weights. Its construction begins with the addition of Property 4:

(4) *Exclusion*: a regressor $X_j$ with $\beta_j = 0$ should receive a zero share.

Unlike the LMG, a regressor whose coefficient is zero receives zero share asymptotically (Grömping, 2007a, p. 10). In practice, the LMG is often computed as though all regressors with zero coefficients have zero shares by excluding them from the model. The distinction between the LMG and PMVD is substantial, both conceptually and numerically.

The PMVD for variable $X_i$ is calculated as:

$$PMVD(X_i) = \frac{1}{n!} \sum_r p(r) seqR^2(\{X_i\}|r)$$

(6)

where the $p(r)$ are the data-dependent weights proportionate to $L(r)$, defined as

$$L(r) = \left[ \prod_{i=1}^{n} seqR^2(\{X_{r_{i+1}}, \dots, X_{r_n}\}|\{X_{r_1}, \dots, X_{r_i}\}) \right]^{-1}$$

The $p(r)$ are just the $L(r)$ normalized to sum to zero:

$$p(r) = L(r) \bigg/ \sum_r L(r)$$

(8)

Comparisons of equations (5) and (8) shows that both LMG and PMVD are averages of the sequential increases in $R^2$, with the LMG putting equal weight on all orders and the PMVD putting variable weights on the orders.

Both the LMG and PMVD have additional properties[3]

(5) *Anonymity*: variable importance is not affected by the labels/positions of the regressors.

(6) *For numbered lists, variable importance does not depend on anything but the first two moments of the joint distribution of the variables.* This is satisfied by the LMG and the PMVD because only the variance-covariance matrix is needed to compute them.

(7) *The addition of a pure noise variable, independent of y and $\{X_1, \ldots, X_n\}$, to a subset of variables does not affect the importance of the subset relative to the other variables.*

(8) *Variable importance should balance conditional and marginal considerations.* The metric should balance the contribution of $X_j$ when alone in the model (direct effect), the contribution of $X_j$ in addition to all other regressors (total effect) and the contributions of $X_j$ considering different subsets of further regressors.

(9) *Orthogonal compatibility: The decomposition respects orthogonal subgroups, i.e., for each orthogonal subgroup of regressors, the assigned shares sum to the unique overall* *model variance (or $R^2$) for that subgroup*. The LMG and PMVD satisfy this by construction.

(10) *Non-negativity: all allocated shares are always non-negative*. The LMG and PMVD satisfy this by construction.

Both the LMG and PMVD can be computed for data with weights or missing values (Grömping, 2018, pp. 10-17).

Additionally, Thomas, Kwan and Zumbo (2018, p. 7) prove for the LMG

(11) *The nonsingular linear transformation of a subset $\{X_1, \dots, X_q\}$ into the subset $\{X'_1, \dots, X'_q\}$ does not affect its importance relative to the other variables*.

### 2.1 Grouping Variables

When some regressors are multichotomous dummy variables or the analyst wishes to group some variables, both the LMG and PMVD can be computing by treating the group as a single variable. In the case of the LMG, this is equivalent to computing the Owen (1972) Value. Variance-covariance matrices can be used to compute these with the mild complication that computations have to be adjusted to account for the groupings.

Grouping multichotomous dummies is both natural and necessary. Dummy variables represent either nominal or ordinal variables or binned values of a continuous variable, perhaps with interactions included. Suppose that a set of two dummies represent a variable $X_i$ with the levels {"Low","Medium","High"}. Assume that "Low" is taken to be the baseline. It is nonsensical to measure the importance of the regressors representing the values of "Medium" and "High" when the variable of interest is $X_j$, of which the dummies represent values.

Measuring the importance of those regressors taken together is how the importance of $X_j$ is measured.

Grömping (2007a) suggests grouping regressors for computational efficiency. This is motivated by models used in marketing that often have 30 regressors. These groupings reduce the number of covariances that need to be computed. The tradeoff is that variance importance is computed with less accuracy. It is unclear how much effect this has on conclusions drawn from these models for ungrouped regressors.

## 2.1  Bootstrapping Confidence Intervals for the LMG and PMVD

Confidence intervals for both the LMG and PMVD can be obtained from the bootstrap. Feldman (2005, p.10) proposes bootstrapping the PMVD using a fixed design and i.i.d. normal errors. Grömping (2007a) implements several methods using R's (R Development Core Team 2008) boot package and recommends the BCa bootstrap. Grömping (2007a, 2007b) finds that the confidence intervals generated by the bootstrap are too wide, with the proviso that the BCa bootstrap wasn't tested due to its long run time. Grömping (2007b) shows, for a particular example with correlated regressors, that the PMVD's confidence intervals are wider than those of the LMG. Strobl (2009) finds a similar result for correlated regressors, but the exact opposite for uncorrelated ones. These results are in keeping with the causal and predictive bases of the LMG and PMVD, respectively. They produce narrower confidence intervals for the types of regressors they prefer: uncorrelated for the PMVD and correlated for the LMG. These results have the proviso that, in addition to the confidence intervals' being too wide, the exact type of bootstrap is not specified.

## 2.2 Comparison of the LMG and PMVD

The fundamental distinction between the LMG and PMVD is that the LMG estimates the regressors' marginal contributions while the PMVD is more conditional.[4] The LMG is equivalent to the Shapley value, which is interpreted to mean the marginal value of each player or regressor in this context. The LMG is compatible with any causal structure among the regressors, including those with zero coefficients. See Grömping (2009), Figure 1 and its discussion in for an illustration of how a regressor with coefficient 0 can still be assigned importance though its causation of another regressor. The PMVD's exclusion criterion rules out this type of causal structure: every regressor must have a direct causal effect on the dependent variable. So, while the LMG has a marginal interpretation, the PMVD is closer to a conditional one (Grömping, 2015, p. 144). The LMG is useful for estimating the importance of all causal variables and is thus explanatory. The PMVD is predictive by focusing on those variables that predict the dependent variable: regressors with zero coefficients have no predictive power. The LMG's focus on explanation allows for more model uncertainty by admitting more causal structures. As the regressors approach perfect correlation, the LMG assigns an equal share to each regressor, while PMVD's shares are data-dependent. Thus, the LMG is agnostic with regard to regressors' importance in the case of perfect multicollinearity, while the PMVD is influenced by the regressors' scales. However, from a predictive point of view, correlated regressors reduce MSE (Abdullaev and Geidarov, 1985, pp. 67-68), so that a model containing them may have its MSE increased by removing one or more correlated regressors. A final important difference is that software to implement the PMVD is patented in the U.S. while the LMG is in the public domain. Table 1 summarizes these differences.

Table 1. Comparisons between LMG and PMVD.

| Criterion | | LMG | PMVD |
|---|---|---|---|
| Marginal or conditional? | | Marginal | Conditional |
| Interpretation | | Explanation | Prediction |
| Compatible with all causal structures that include the regressors? | | Yes | No |
| Accounts for model uncertainty? | | More | Less |
| Confidence intervals | Correlated regressors | Narrow | Wide |
| | Uncorrelated regressors | Wide | Narrow |
| Limit of $R^2$ shares as correlations between regressors converge to 1. | | $\frac{1}{n}$ | Data-dependent |
| Availability | | Public domain | Patented in U.S. |
| Approximations available without grouping variables?[5] | | Yes | No |

### 2.3 Software for Variable Importance in Linear Models

Variable importance has been implemented in both R and SAS. Grömping's (2007a) R package relaimpo computes several variable importance measures including LMG and, outside the U.S., PMVD. Package relaimpo uses the variance-covariance matrix for efficiency, allows groupings and two-way interactions, and can obtain bootstrapped confidence intervals. Coleman's (2017) SAS macro %DECOMPOSE_R2 runs slower regressions to estimate the LMG with up to $n$-way interactions, with some restrictions. %DECOMPOSE_R2 does not estimate confidence intervals, but can be used inside a bootstrap loop to estimate them.

## *2.4 Large numbers of regressors*

When the number of regressors becomes too large, computing the LMG and PMVD becomes computationally infeasible. This problem is worsened if confidence intervals are required. The LMG requires computing $2^n$ values of $seqR^2$.[6] The storage requirement scales similarly. In these cases, sampling (van Campen et al., 2018) or linear approximation (Fatima et al., 2008) algorithms can approximate the LMG.[7] The PMVD is even more complex, requiring computing $seqR^2$ for all $n!$ permutations of the regressors, then using these computations to create weights, so that it runs in factorial time and space. However, recursion (9) requires computing only $2^n$ values of $R^2$:

$$P(S) = R_S^2 \left( \sum_{j \in S} P(S \backslash j)^{-1} \right)^{-1}$$

$$(9)$$

for $S \subset N$ with $P(\emptyset) = c > 0$. Then, $PMVD_j = P(N)/P(N \backslash j)$. To see that this requires only $2^n$ evaluations of $R^2$, note that when computing PMVD for all $j$, the first stage requires only the overall $R^2$, the second requires each value of $R^2$ without one regressor, the second stage deletes 2 regressors, and so on until the last stage deletes $n - 1$ regressors, that is, it only uses one regressor in each regression. Thus, the total number of $R^2$ evaluations is $\sum_{k=1}^{n} \binom{n}{k} = 2^n$. This cannot be reduced further, unlike the LMG, because deleting any one evaluation makes calculating one or more $PMVD_j$ infeasible. Moreover, it is unclear whether any approximation scheme is feasible given the dependencies the recursion induces.

Grömping (2018) allows the user to group regressors to reduce run time. As shown in Section 3, this can become perilously close to using *p*-values to measure "oomph."

Grömping (2015, pp. 146, 2016) suggests using the Genizi (1993)/Johnson (2000) metric to approximate the LMG in large samples. This is based on the numerical example in Grömping (2015, Table 2, p. 147) and Johnson's (2000) observations. Johnson (2000) provides a simple way to compute this metric. The behavior of the Genizi/Johnson metric has to be investigated more fully before this suggestion can be adopted.

### 2.5 Variable Importance in Ensemble Models

Ensemble models like random forests build large numbers of regression trees to model input data. Random forests repeatedly select a subset of observations with replacement to build a single tree. Breiman's (2001) procedure uses CART with one of a preselected number of randomly selected variables to build the trees (RF-CART).[8] Hothorn et al. (2006) use conditional inference to build the trees (RF-CI). Thus, the choice of method determines whether the marginal or conditional view of variable importance should be used. While Hothorn et al. (2006) were motivated by RF-CART's biases, Altmann et al. (2010) propose PIMP-RF, an improvement to RF-CART to remove its biases while preserving its other properties.

Random forests have many variable importance measures. This paper focuses on MSE-reduction, a variant of Breiman's (2002) Method 1. For each tree $t$, the out-of-bag observations, that is, those observations not included in $t$, are selected. Then, OOB-MSE, the mean squared error for all out-of-bag (OOB) observations is calculated as

$$\text{OOB-MSE} = \frac{1}{n} \sum_{i=1}^{n} \left( y_i - \overline{\widehat{y_{i\text{OOB}}}} \right)^2$$

(10)

where $\overline{\widehat{y_{i\text{OOB}}}}$ denotes the average prediction for the $i$th observation from all trees for which this observation has been OOB. Analogously to linear regression, with the overall sum of squares $\text{SST} = \sum_i^N (y_i - \overline{y})^2$ defined in the usual way, OOB-$R^2$ can be obtained as $1 - \text{OOB-MSE/SST}$. (Grömping, 2009, p. 311)

Then, for each tree, the average squared prediction error of the OOB observations, $\text{OOBMSE}_t$, is calculated. Next, each regressor $X_j$ is randomly permuted. The average squared error of the prediction using the permuted version of $X_j$, $\text{OOBMSEP}_{j,t}$, is computed. MSE reduction for regressor $X_j$ is defined as the average of the difference $\text{OOBMSEP}_{j,t} - \text{OOBMSE}_t$ over all trees, taking this difference to be 0 whenever $X_j$ is not included in tree $t$. MSE reduction can be negative. Its sum needs not equal OOB-MSE, so each regressor's relative share can be used as its variable importance measure. RF-CI uses a slightly more complicated permutation scheme (Strobl et al., 2008) to obtain the $\text{OOBMSEP}_{j,t}$. MSE reduction associated with a variable $X_j$ is not the same as the MSE reduction due to its presence (Ishwaran et al., 2008). This property enables the use of MSE reduction as a variable importance measure.

Strobl et al. (2008) provide theoretical evidence that RF-CART prefers correlated regressors while RF-CI prefers uncorrelated regressors. This is consistent with Abdullaev's and Geidarov's (1985, pp. 67-68) proof regrading linear models. RF-CART is thus more focused on explanation: correlated regressors can reflect indirect causality. RF-CI, by ruling out these regressors, focuses on prediction. Section 7 shows how RF-CART and RF-CI can be combined to yield at least a partial description of causality.

Ishwaran et al. (2008) develop methods to generate confidence intervals for random forest variable importance methods. These include two fast subsampling methods. Ishwaran and Kogular (2019) implement them in R package randomForestSRC for Breiman's (2002) Method 1, which measures OOB accuracy improvement.

## 3 Variable Importance as "Oomph"

Pogrow (2019, p. 229-230) suggests using $seqR^2$ computed for one variable in the full model as a measure of "oomph" in nonexperimental research. He further suggests using 15% as a cutoff to determine whether a variable is sufficiently "oomphy." Pogrow is fully cognizant of potential pitfalls: $p$-hacking, lack of relevant research and small samples masking potentially useful research. Pogrow claims that $p$-hacking is mitigated, though this is disproved for a very literal interpretation of the cutoff criterion below. The LMV and PMVD are methods to measure "oomph."[9]

Taking Pogrow's (2019) suggested use of a cutoff very literally leads one back to using $p$-values. This concept is called "usefulness." Theil (1971, pp. 168-181) rediscovered it as "incremental contribution." Theil (1971, p. 175), Darlington (1968, pp. 168-169) and Bring (1996, p. 59) show that it is equivalent to using $t^2$. From Theil (1972, p. 175),

$$t_j^2 = \frac{R^2 - R_{-j}^2}{1 - R^2} \frac{1}{m - n}$$

$$(11)$$

where $R_{-j}^2$ is the $R^2$ from the regression omitting regressor $X_j$ and $m$ is the number of observations.  Maximizing the sum of "usefulness" of one or more variables is equivalent to minimizing their two-sided $p$-values, that is, $p$-hacking.

Instead, the cutoff should be restated in terms of the LMV or PMVD.  This avoids $p$-values by using all of the correlations between the regressors.  Since the criterion's values are stochastic, confidence intervals should be considered.  The exact choice of cutoff or confidence intervals is likely to be subject-specific and beyond the scope of the present article.  Pogrow (2019) is concerned with prediction, so the PMVD is a natural choice.  The LMG is more natural for selecting variables for intervention or to understand the causal structure.  It may require fewer resources to intervene in a nonpredictive variable that causes a nonzero regressor than for the regressor itself.  If direct intervention in the regressor is physically or legally impossible, then intervening in one of its causes, if possible, is the only way to change it.  Interventions are explored more fully in Section 6.  The LMG is also in the public domain everywhere unlike the PMVD.

A further implication is that variable importance can replace or supplement $p$-values in reporting, with the proviso that their confidence intervals should also be reported.  As Ziliak and McCloskey (2008, 12-13, 23-27) note, a $p$-value is a measure of the precision with which a regressor's coefficient is estimated.  A negligible regressor is always negligible regardless of how precisely it has been estimated.  Variable importance identifies negligible regressors by assigning very little importance to them.  A negligible regressor has insufficient "oomph" by definition.  Its statistical significance is irrelevant.

By construction, the LMG and PMVD measure the importance of a regressor with a nonzero coefficient relative to the same regressor with a zero coefficient.  Often, the analyst is interested in the difference of one or more coefficients from nonzero constants.  Variable importance can be employed by transforming the dependent variable so that the departures of the coefficients of the regressors from zero measure the differences between them and the constants of interest. We show this for one regressor. Suppose the analyst wishes to use variable importance for $\beta_1 = C$, or, equivalently, $\beta_1 - C = 0$ in the regression

$$\boldsymbol{y} = \beta_0 + \beta_1 X_1.$$

$$(12)$$

Let $\boldsymbol{y}^* = \boldsymbol{y} + CX_1$. Then regression (11) becomes

$$\boldsymbol{y}^* = \beta_0 + \beta_1^* X_1$$

$$(13)$$

where $\beta_1^* = 0$ implies $\beta_1 = C$. Variable importance can then measure the "oomph" of a departure of $\beta_1$ from $C$. Again, statistical significance is irrelevant when the departure lacks sufficient "oomph." This generalizes to any number of individual regressors. No similar transformation exists for linear sums of regressors.

It is not clear how to measure "oomph" in random forests, given that the variable importances do not sum to an $R^2$-like value. One possibility is to multiply the variable importance shares by $\text{OOB-}R^2$. Confidence intervals can be obtained by multiplying the confidence bounds by $\text{OOB-}R^2 / \sum_{i=1}^n \text{VI}_i$ where the denominator is the sum of the unadjusted variable importances.

## 4 Variable Importance as a Modelling Guide

The idea of using variable importance to select regressors has a long history. Using $p$-values as a selection criterion is (still) standard practice. In effect, the lower the $p$-value, the more importance the regressor is given.  Stepwise regression is notorious for this.  R's caret package does this in linear models using the equivalent *t*-statistic.  In contrast, De Veaux (2017, p. 152) recommends using ensemble models to find variables with greatest "contribution" to model fit, that is, importance, for further study.  Grömping (2009) finds that RF-CART produces variable importances similar to the LMG and those of RF-CI are similar to the PMVD.  It is very possible for both methods to select the same variables.

In the linear model context, LMG and PMVD should, with few exceptions, only be used on final models to avoid importance-hacking.  De Veaux's and similar two-stage methods can be used to select and analyse variables.  The key is that one method selects the variables for analysis by another, unrelated method.  Any attempt to manipulate variable selection will have unpredictable results on the final model.  The use of LMG or PMVD alone to reduce the number of regressors in a final model carries the risks of bias and variance increase in addition to importance-hacking.

In random forests, discarding unimportant variables has been done since Breiman (2001) and is considered to be a feature.  The risk of importance-hacking vanishes because all possible regressors should be included when building the model.  Removing unimportant regressors does little damage to fit while reducing the number of regressors below the number of observations when necessary.

## 5 Discerning Causal Structure

Measuring variable importance can be the first step to discerning at least a partial causal structure.  Analyst evaluation of the plausibility of causal links is required.  This section assumes that every potential causal link is plausible and that the correlations between all variables are low enough to avoid confounding.  Moreover, the indirectly causal variables are assumed not to be mediators or suppressors. The analyst can use MacKinnon et al.'s (2000) unified statistical framework for mediation, confounding and suppression to investigate. Suppose that RF-CART finds that regressors $X_1$, $X_2$ and $X_3$ are important while RF-CI only selects $X_1$ and $X_2$. This has the ready interpretation that $X_3$ indirectly causes the outcome $y$. Figure 1 shows the 3 possible causal paths. The obvious method to choose among these is to look at correlations. This is complicated by the 9 combinations of nominal, ordinal and continuous variables for each pair of regressors of interest. Assuming the complications are worked out or the regressors are all of the same type, the pairs whose correlations have sufficient "oomph" are accepted as causal. This generalizes to multiple indirectly causal variables with the proviso that hierarchies among indirectly causal variables cannot be distinguished. This procedure is amenable to automation. Figure 2 illustrates three causal structures compatible with variables $X_2$ and $X_3$ causing directly causal variable $X_1$. Only if $X_2$ and $X_3$ are sufficiently uncorrelated with each other can one of the structures, (c), be selected. Thus, this procedure can sometimes deduce only partial causal structures.

Figure 1. The three causal structures compatible with $X_1$ and $X_2$ directly causing $y$ and $X_3$ indirectly causing $y$. (a) $X_3$ causes $X_1$. (b) $X_3$ causes $X_2$. (c) $X_3$ causes both $X_1$ and $X_2$.

(a)

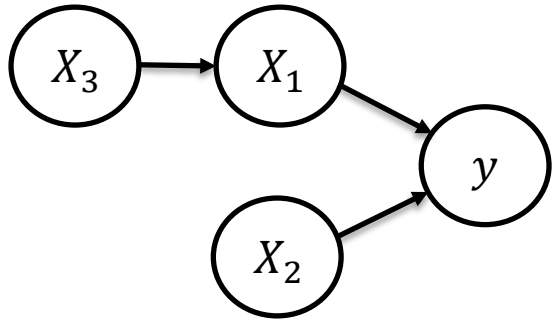

(b)

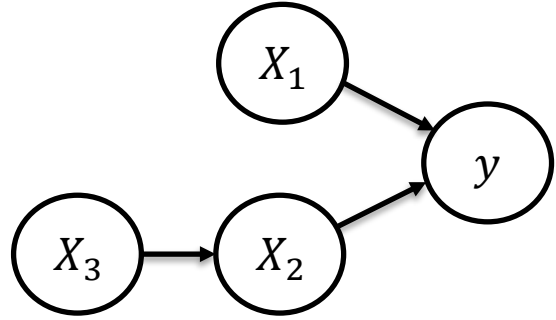

(c)

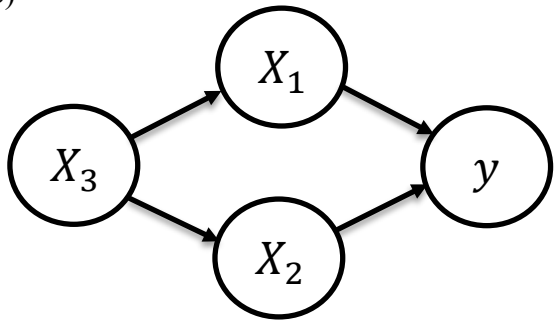

Figure 2. The three causal structures compatible with $X_1$ and $X_2$ directly causing $y$ and $X_2$ and $X_3$ indirectly causing $y$. (a) $X_3$ causes $X_2$. (b) $X_2$ causes $X_3$. (c) $X_2$ and $X_3$ both cause $X_1$.

(a)

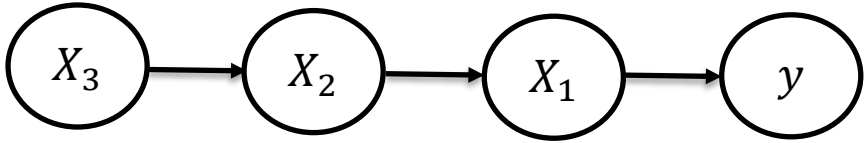

(b)

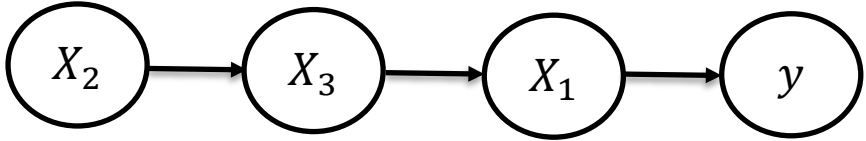

(c)

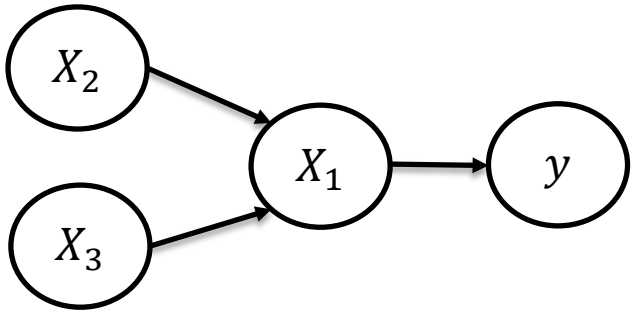

Causal information about regressors with limited predictive value cannot be obtained. Suppose that, additionally, RF-CART identifies regressors $X_4$ and $X_5$ as causal while RF-CI rejects them. Further, suppose that $X_4$ and $X_5$ are correlated both with each other and with $X_1$ or $X_2$ and that the analyst rejects their causality of $X_1$ and $X_2$. This situation can result from RF-CI's emphasis on prediction rejecting correlated regressors to reduce MSE. In this case, no causal statement about $X_4$ and $X_5$ can be made other than they directly or indirectly cause $y$.

While the extension to linear models is theoretically straightforward, in practice, it may be infeasible. When fitting linear models, variables with zero coefficients are normally excluded. However, a theoretically zero coefficient variable may have "oomph," thus lending its model to similar causal interpretation as for random forests using LMG and PMVD. It is not clear whether and how often this occurs in practice, so this remains theoretical speculation.

## 6 Variable Importance as a Guide to Interventions

Variable importance can be used to guide interventions. Variables are chosen for intervention in order of importance. Fields (2003) reports success using a Shapley value-like decomposition of the variance of the dependent variable (Israeli, 2007, pp. 202-203) which produces similar importance rankings. Fields (2003, p. 16) also notes that interventions may be subject to legal limitations. Physical limitations can be added to this. The LMG and other causal measures are to be preferred because intervention is based on causality. Interventions can be made to indirectly causal variables which the PMVD and other predictive measures ignore. Grömping (2007b, p. 146) echoes other authors' concerns that it is better to know the full causal structure when designing interventions. Section 5 has shown how it may be possible to extract the causal structure from the data using variable importance. It is only when causal structure information is completely lacking that variable importance should be used alone. Grömping further notes that using variable importance is not without its risks: an intervention can change the regressors' correlation structure so as to reduce its effect on the dependent variable below that which would be expected from the original correlation structure.

## 7 Conclusions and Extensions

Variable importance has a wide variety of applications beyond measuring or ranking variables' contribution to model fit. Depending on whether the analysis is explanatory or predictive, different methods to obtain variable importance are appropriate. A variable's importance can measure its practical effect, or "oomph," be used in developing models or designing interventions. Explanatory and predictive measures of variable importance can be combined to at least partially deduce the causal relationships between variables. The measures for linear models in the present paper are based on dispersion, while the measures for ensemble models are based on both level and dispersion via MSE. It is possible that contexts not considered may require other constructs.

In the linear context, the LMG is recommended for estimating causal importance and the PMVD is recommended for predictive importance. Both of these are computationally and storage intensive, especially the PMVD. The LMG can be approximated in these situations using sampling, linear approximation and, possibly, the Genizi/Johnson decomposition. Approximation methods for the PMVD are unknown and probably impossible. An important area of future research is to determine whether these methods are possible, then to develop them. The currently available bootstrap confidence intervals are too wide except, perhaps, BCa bootstrap intervals. Given the recent increases in computing power, it should be possible to thoroughly investigate all these approximation and confidence interval methods.

Random forests have two recommended variants, PIMP-RF and RF-CI, for causal and predictive use, respectively. Unlike linear models, variable importance is naturally captured during tree construction. This often comes at the cost of model interpretability. Random forests can be used as a screening tool to select variables for further analysis by other methods, such as linear regression. This has the virtue of focusing on important variables and avoiding importance-hacking of the final model.

Hierarchical partitioning can be used for models that lack specialized variable importance estimation methods. Currently, hierarchical partitioning is explanatory. It should be possible to extend it to predictive modelling, with a similar penalty as in linear models in computational and storage costs. Approximation methods have yet to be developed. In some cases, these may be impossible due to model complexity. Bootstrapped confidence intervals are available. In every case, a version can be obtained by using a fixed design and randomized errors.

**Footnotes**

1. The text for Property 1 is modified from Grömping (2015). Properties 2-4 are copied from Grömping (2015).
2. For fuller details about the computations, see Grömping (2007a).
3. Properties 5-11 are from Grömping (2015) with modified descriptions.
4. See Shmueli (2010) for an excellent discussion of explanation and prediction.
5. This is covered in Subsection 2.4 below.
6. This can be seen by simply noting that every subset of regressors can be generated by using a binary vector of length $n$ to select regressors. The number of these vectors is $2^n$.
7. LMG is in computational complexity classs #P-complete (Deng and Papadimitriou, 1994), essentially meaning that fully polynomial-time randomized approximation schemes (FPRAS) may exist to any arbitrary precision.
8. "RF-CART" and "RF-CI" are Grömping's (2007) abbreviations.
9. This discussion does not take into account instances when "oomph" is better measured by a level measure, such as the purchasing power parity example of Ziliak and McCloskey (2008, pp. 94-97). It is an empirical matter to determine the relationship of nonlevel variable importance to the "oomph" of a level change.